\documentclass[aps,twocolumn]{revtex4}
\usepackage{epsfig}

\begin{document}
\renewcommand{\thefootnote}{\fnsymbol{footnote}}
\renewcommand{\theequation}{\arabic{section}.\arabic{equation}}
\begin{widetext}
\begin{small}
A. Bonanni and T. Dietl, A story of high-temperature ferromagnetism in semiconductors, {\em Chem. Soc. Rev.} {\bf 39} (2010) 528-539, http://dx.doi.org/10.1039/B905352M --– Reproduced by permission of The Royal Society of Chemistry.
\end{small}
\end{widetext}

\ \\

\title{A story of high-temperature ferromagnetism in semiconductors}

\author{Alberta Bonanni$^a$ and Tomasz Dietl$^{b,c}$}
\email[]{E-mail: Alberta.Bonanni@jku.at, Dietl@ifpan.edu.pl}

\affiliation{$^a$ Institute for Semiconductor and Solid State Physics, Johannes Kepler University, A-4040 Linz, Austria\\
$^b$ Institute of Physics, Polish Academy of Sciences, PL-02-668 Warszawa, Poland\\
$^c$ Institute of Theoretical Physics, University of Warsaw, PL-00-681 Warszawa, Poland}

\date{\today}

\begin{abstract}
\noindent The comprehensive search for multifunctional materials has resulted in the discovery of semiconductors and oxides showing ferromagnetic features persisting to the room temperature. In this tutorial review the methods of synthesis of these materials systems, as well as the application of element-specific nano-analytic tools, particularly involving synchrotron radiation and electron microscopy, are described and shown to reveal the presence of nano-scale phase separations. Various means to control the aggregation of magnetic cations are discussed together with the mechanisms accounting for ferromagnetism of either condensed or diluted magnetic semiconductors. Finally, the question of whether high temperature ferromagnetism is possible in semiconductors not containing  magnetic ions is touched upon.

\end{abstract}

\maketitle

\section{Introduction}
\label{intro}
Today's epitaxy of semiconductors, in addition to represent a major tool to synthesise complex high quality multi-layer structures, is also successfully exploited to fabricate self-organised quantum dots, nanowires, and nanocolumns. This variety of quantum nanostructures allows one to explore novel physics and device concepts in semiconductor systems with reduced dimensionality, expected to form the building blocks of future sensor, information, and communication technologies. At the same time, the continuous progress in the development of magnetic media, multilayers, and nanopillars of ferromagnetic metals, in which not only the charge transport senses the magnetisation orientation, but the magnetisation orientation can be controlled by electric current, offers scenarios of fast and non-volatile universal memory and low-power hybrid or all-magnetic logic.

In this {\em tutorial review} we describe the recent progress in the understanding and control of {\em epitaxially grown tetrahedrally coordinated} semiconductor films containing transition metals (TMs), typical examples being (Cd,Mn)Te, (Zn,Co)O, (Ga,Mn)As, (Ga,Fe)N, and (Ge,Mn). Since the discovery of ferromagnetism in (In,Mn)As and (Ga,Mn)As two decades ago \cite{Ohno:1998_a} and then in a broad spectrum of other diluted magnetic semiconductors (DMSs) and diluted magnetic (DMOs) \cite{Liu:2005_a,Chambers:2006_a}, these systems have attracted a considerable attention due to their potential to bridge the above-mentioned resources of functional semiconductors and ferromagnetic metals.

Owing to the use of advanced element-specific nanocharacterisation tools it becomes increasingly clear that ferromagnetic DMSs form two distinct classes. To the first class belong materials, in which the presence of robust ferromagnetism correlates with the existence of nanoscale regions containing a large density of magnetic cations, that is with the formation of {\em condensed} magnetic semiconductors (CMSs) buried in the host matrix and specified by a high spin ordering temperature. Interestingly, the aggregation of CMSs and, therefore, the ferromagnetism of the resulting composite system shows a dramatic dependence on the growth conditions and co-doping with shallow impurities. Hence, as we emphasise in this survey, the epitaxy of magnetically doped semiconductors constitutes a versatile way of fabricating in a self-organised fashion semiconductor/ferromagnet nanocomposites with hitherto unexplored but striking functionalities at ambient temperature.

The second class of ferromagnetic DMSs comprises $p$-type semiconductors containing  {\em randomly} distributed TM impurities. In these materials --- as underlined in this review --- holes residing in the valence band, rather than carriers in impurity states or conduction band electrons, can mediate efficient long-range ferromagnetic spin-spin couplings. We list a number of outstanding low-temperature functionalities demonstrated recently for structures containing hole-controlled ferromagnets, and being now successfully transferred to ferromagnetic metal systems. We describe also the upshots of a long-standing effort to increase the TM and hole densities in functional semiconductors, in defiance of solubility limit, self-compensation, and hole binding by TM impurities, rather early identified \cite{Dietl:2000_a} as the main obstacles to the still uncovered way towards a successful synthesis of a room temperature spatially uniform ferromagnetic semiconductor.

Over the last decade, the field of DMSs has evolved into an important branch of condensed matter physics and materials science, whose bibliography accounts for over 7,000 publications. In the spirit of tutorial reviews we quote here only a few original papers, addressing the readers to a series of specific review articles and book chapters, in which comprehensive lists of references to research papers can be found.

\section{Incorporating magnetic ions into semiconductors}
It is well known that the phase diagrams of a number of semiconductor alloys exhibit a miscibility gap in a certain concentration range. For instance, in the case of (Ga,In)N, at high In concentrations the alloy decomposes into In-rich quantum-dot like regions embedded in the In-poor matrix. Particularly low is the solubility of TM impurities in tetrahedrally coordinated semiconductors, so that low-temperature epitaxy or ion implantation have to be employed to introduce a sizable amount of the magnetic constituent. An exception here is the large solubility of Mn in II-VI compounds, often exceeding 50\%, so that in the diluted case the Mn atoms remain distributed randomly over the substitutional cation sites \cite{Pajaczkowska:1978_a,Furdyna:1988_a}, even if the alloy is grown close to thermal equilibrium, as in the case of, {\em e.g.}, the Bridgman method.

The large solubility of Mn in II-VI compounds can be associated to the fact that the Mn $d$ states little perturb the $sp^3$ tetrahedral bonds as both the lower $d^5$ (donor) and the upper $d^6$ (acceptor) Hubbard levels are respectively well below and above the band edges \cite{Zunger:1986_a,Dietl:1994_a}. This qualitative picture is supported by first principles computations, showing the virtual absence of an energy change associated with bringing two Zn-substitutional Mn atoms to the nearest neighbor cation sites in (Zn,Mn)Te, $E_{\mathrm{d}} = 21$~meV \cite{Kuroda:2007_a}. In contrast, according to a pioneering {\em ab initio} work \cite{Schilfgaarde:2001_a} and to the subsequent developments \cite{Kuroda:2007_a,Katayama-Yoshida:2007_a}, a particularly strong tendency to form non-random alloys occurs in the case of DMSs in which the $d$ orbitals are close to the Fermi energy and thus contribute significantly to the bonding. For instance, the pairing energy of two Ga-substitutional Mn atoms is $E_{\mathrm{d}} = -120$~meV in GaAs and $-300$~meV in GaN, and reaches $-140$ and $-350$~meV in the case of a cation-substitutional Cr nearest neighbor pair in ZnTe and GaN, respectively \cite{Schilfgaarde:2001_a,Kuroda:2007_a}.

The rapid progress in the DMS research that started in the 1990's has stemmed, to a large extent, from the development of methods enabling material synthesis far from thermal equilibrium, primarily molecular beam epitaxy (MBE), but also pulse laser deposition (PLD) and then metalorganic vapor phase epitaxy (MOVPE), atomic layer deposition (ALD), and sputtering. These methods have made it possible to obtain DMS films with a concentration of the magnetic constituent beyond the solubility limits at thermal equilibrium. Similarly, doping during an epitaxy process allows one to substantially increase the electrical activity of shallow impurities. In the case of III-V compounds, in which divalent magnetic atoms supply both spins and holes in the valence band, the use of  low-temperature molecular beam epitaxy (LT MBE) provides thin films of, {\em e.g.}, Ga$_{1-x}$Mn$_x$As with a Ga-substitutional Mn concentration $x$ up to 0.08 and with a hole concentration in excess to $10^{20}$~cm$^{-3}$ \cite{Matsukura:2008_a,Jungwirth:2008_a}.

Another important material issue in DMSs is the existence of an upper limit for the achievable carrier density under thermal equilibrium conditions in particular hosts. The existence of such a limit -- a well known challenge in the development of semiconductor laser diodes -- precludes in the context of DMSs the achievement of an arbitrary strong carrier-mediated coupling between diluted localised spins. The limited carrier doping efficiency may originate from the finite solubility of a particular dopant in a given host.  In most cases, however, the effect of self-compensation --- consisting in the appearance of compensating point defects once the Fermi level reaches an appropriately high position in the conduction band (donor doping) or low energy in the valence band (acceptor doping) --- is involved.

According to particle-induced x-ray emission (PIXE) \cite{Furdyna:2008_a} measurements an attempt to increase the Mn or Be acceptor concentration in (Ga,Mn)As not only results in the formation of MnAs precipitates but also in the occupation by Mn of {\em interstitial} positions, Mn$_\mathrm{I}$. Since Mn$_\mathrm{I}$ has two unbound $4s$ electrons, it acts as a double donor in GaAs \cite{Jungwirth:2006_a}. Its formation is triggered by a lowering of the system energy due to the removal of holes from the Fermi level. Moreover, a symmetry analysis demonstrates that the spin-dependent coupling between the holes and the Mn${_\mathrm{I}}$ ions is weak and that the exchange interaction between Mn${_\mathrm{I}}$ and Mn${_\mathrm{Ga}}$ is antiferromagnetic \cite{Jungwirth:2006_a,Furdyna:2008_a}.  The resulting detrimental effect of interstitials on the ferromagnetism can be partly reduced by an annealing process that promotes the diffusion of the Mn${_\mathrm{I}}$ ions to the surface, where they partake in the formation of an antiferromagnetic MnO layer \cite{MacDonald:2005_a}. This post-growth thermal treatment leads to a substantial increase in the magnitude of both $T_{\mathrm{C}}$ (currently reaching up to 180~K \cite{Olejnik:2008_a}) and the spontaneous magnetisation. However, since at high hole concentrations the formation energies and diffusion barriers of all defects tend to decrease, further work is necessary to prove that the interstitials are the only relevant compensating centers in (Ga,Mn)As and other $p$-type DMSs.

\section{Nanoscale phase separations and high-temperature ferromagnetism}

As mentioned above, with the exception of Mn in II-VI chalcogenides and oxides, the solubility of TM impurities in tetrahedrally coordinated semiconductors is rather low at thermal equilibrium conditions. In particular, as reviewed elsewhere \cite{Dietl:2007_a,Tanaka:2008_a}, both hexagonal and zinc-blende Mn-rich (Ga,Mn)As nanocrystals are clearly seen by high-resolution transmission electron microscopy (HRTEM) in (Ga,Mn)As either annealed or grown at appropriately high temperatures. In this case the formation of hexagonal inclusions is the result of crystallographic decomposition, while the cubic nanocrystals are the evidence of chemical phase separation, this last known in the DMS literature as spinodal decomposition, independently of the microscopic mechanism leading to the aggregation of the TM cations. Magnetisation measurements by highly sensitive magnetometry involving superconductive quantum interferometer device (SQUID), point to an apparent Curie temperature of 320~K and 360~K for these two kinds of nanocomposite systems, respectively.

\begin{figure}[htb]
\includegraphics[width=0.9\linewidth]{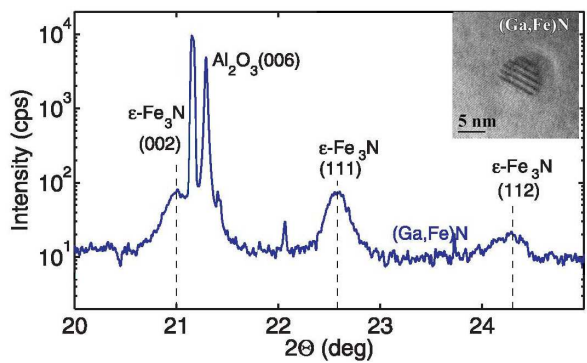}
\caption{Synchrotron x-ray diffraction and transmission electron microscopy of (Ga,Fe)N showing the precipitation (crystallographic phase separation) of hexagonal $\epsilon$-Fe$_3$N nanocrystals. The diffraction peaks detected between in the range 21.2--22.2$^{\circ}$ originate from the substrate and their intensity depends on the tilt of the sample, while the signals from the nanocrystals are virtually tilt-independent (after \cite{Bonanni:2008_a}).}
\label{fig:GaFeN_XRD}
\end{figure}

As emphasised by the present authors \cite{Dietl:2007_a,Bonanni:2007_b}, the detection of the phase separation has been highly challenging in DMSs research. Only recently the actual spatial distribution of TM cations in some DMSs has been established by various groups, owing to the application of state-of-the-art element-specific nanocharacterisation tools. Taking (Ga,Fe)N as an example, we note that according to standard laboratory high-resolution x-ray diffraction (HRXRD), the incorporation of Fe simply leads to a broadening of the GaN-related diffraction maxima without revealing any secondary phases \cite{Bonanni:2007_b}. In contrast, a much brighter synchrotron source has allowed to identify the presence of Fe$_3$N precipitates in the same samples, as shown in Fig.~1, these results being supported by our groups' recent near-edge x-ray absorption fine-structure (EXAFS) studies. At the same time, HRTEM with appropriate mass and strain contrast as well as electron dispersive spectroscopy (EDS), not only corroborated the outcome of synchrotron XDR, but revealed the presence of spinodal decomposition in this system \cite{Bonanni:2008_a}, as shown in Fig.~2.

\begin{figure}[htb]
\includegraphics[width=0.5\linewidth]{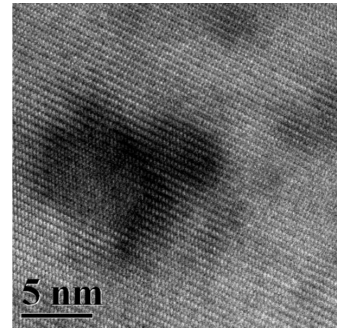}
\caption{Element-specific transmission electron microscopy of (Ga,Fe)N showing chemical phase separation (spinodal decomposition) leading to regions rich in Fe (darker spots) without breaking the continuity of the wurtzite structure (after \cite{Bonanni:2008_a}).}
\label{fig:GaFeN_spinodal}
\end{figure}

The application of TEM with EDS capabilities allowed to evidence the chemical phase separation in (Zn,Cr)Te \cite{Kuroda:2007_a}, (Al,Cr)N, and (Ga,Cr)N \cite{Gu:2005_a}, while according to the results summarised in Fig.~3, coherent hexagonal nanocrystals were detected by spatially resolved x-ray diffraction in (Ga,Mn)N. Finally, we mention the case of (Ge,Mn), where under suitable growth conditions {\em quasi}-periodically arranged nanocolumns are observed by nanoscale chemical analysis using electron energy-loss spectroscopy (EELS), as shown in Fig.~4. Actually, a tendency to nanocolumn formation was also reported for (Al,Cr)N \cite{Gu:2005_a}.  This demonstrates that growth conditions can assist in controlling the nanocrystals shape. Interestingly, these two kinds of nanocrystal forms were reproduced by Monte-Carlo simulations \cite{Katayama-Yoshida:2007_a}.

\begin{figure}[htb]
\begin{center}
\includegraphics[width=0.8\linewidth]{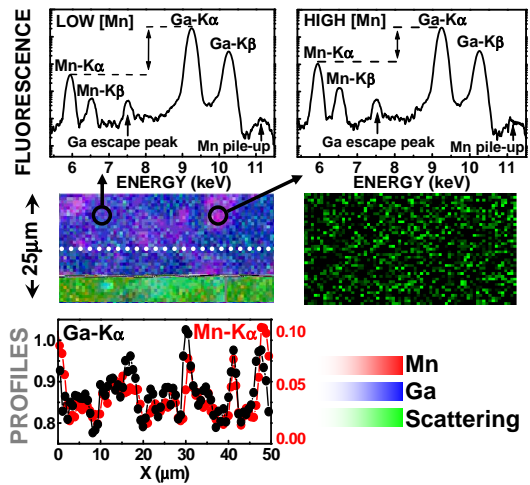}
\caption{Evidence for the formation of Mn rich hexagonal nanocrystals in (Ga,Mn)N from element-specific synchrotron radiation micro-probe (Reprinted with permission from G. Martinez-Criado {\em et al.}, Appl. Phys. Lett., 2005, {\bf 86}, 131927. Copyright 2006, American Institute of Physics).}
\label{fig:GaMnN}
\end{center}
\end{figure}

Since spinodal decomposition-driven magnetic nanocrystals -- that we call condensed magnetic semiconductors (CMSs) -- assume the crystallographic form imposed by the matrix, their properties are not yet included in materials compendia. Accordingly, it is {\em a priori} unknown whether they are metallic or insulating as well as whether they exhibit ferromagnetic, ferrimagnetic or antiferromagnetic spin order. However, due to a large concentration of the magnetic constituent within the CMSs nanocrystals, their spin ordering temperature is expected to be relatively high, typically above the room temperature.

\begin{figure}[htb]
\begin{center}
\includegraphics[width=0.8\linewidth]{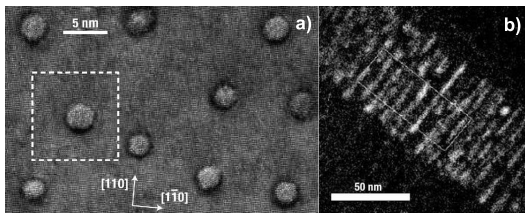}
\caption{Mn-rich nanocolumns in (Ge,Mn) evidenced by: (a) HRTEM plane view and (b) Mn chemical maps derived from EELS measurements [Reprinted by permission from Macmillan Publishers Ltd: Nature Mat.
(M. Jamet {\em et al.}, Nature Mat., 2006, {\bf 5}, 653), copyright (2006)].}
\end{center}
\label{fig:GeMn}
\end{figure}

In addition to the case of annealed (Ga,Mn)As discussed above, a strict correlation between ferromagnetic features and the presence of CMS nanocrystals has been demonstrated for a number of DMSs and diluted magnetic oxides (DMOs), notably for (Ge,Mn) \cite{Jamet:2006_a}, (Zn,Cr)Te \cite{Kuroda:2007_a}, and (Ga,Fe)N \cite{Bonanni:2008_a}.  However, the identification of the dominant microscopic mechanisms leading to robust spin ordering for particular combinations of CMSs and hosts, awaits for detailed experimental and theoretical studies. As an introduction to the relevant physics, we list first the microscopic mechanisms accounting for spin order in solids and then we discuss their relevance to the observed ferromagnetic-like features of CMSs.

Finally, we note that depending on the growth conditions CMS nanocrystals can be distributed randomly or accumulate either at the interface with the buffer or at the film surface. Furthermore, TM impurities may decorate or diffuse along extended defects such as dislocations or grain boundaries. This appears to explain an inverse correlation between samples' quality and the appearance of high $T_{\mathrm{C}}$ ferromagnetism, noted by some authors in the case of oxides. All these observations put the accent on the fact that both lateral and vertical spatial resolution are necessary to detect the nanocrystal presence and to probe their properties.

\section{Physics of exchange in solids}
\subsection{Potential and kinetic exchange interactions}

It is well known that the direct dipole-dipole interaction between magnetic moments is much too weak to explain the typical magnitudes of spin dependent interactions between electrons. It is convenient to single out two spin exchange mechanisms accounting for the magnetic properties of solids, discussed in detail in the context of DMSs in a number of reviews \cite{Dietl:1994_a,Kacman:2001_a,Dietl:2008_e}.

The driving force behind the {\em potential exchange} is the Pauli exclusion principle that precludes two electrons with the same spin to appear simultaneously at the same location. Accordingly, the magnitude of the Coulomb potential energy is lower for such a pair, comparing to the case of two electrons with antiparallel spins. The potential exchange is ferromagnetic and accounts for the Hund's rule, intra-atomic $s$-$d$ exchange interaction, and exchange interactions between spins of carriers occupying the same band.

The {\em kinetic exchange} occurs between two electrons residing at different sites. One of these electrons can visit the site occupied by the other provided that its spin has the orientation matching a relevant empty level. Such a quantum hopping, or in other words hybridisation of states, enlarges the localisation radius and, hence, lowers the electron kinetic energy. The kinetic exchange leads usually, but not always, to an antiferromagnetic interaction between the spin pair in question. In compounds containing transition metals, the $p$-$d$ kinetic exchange couples the spins of carriers occupying anion $p$-like bands with the spins of electrons residing in open $d$ shells of magnetic cations. Typically, in the considered TM compounds the kinetic exchange -- if symmetry allowed -- is much stronger than the potential exchange. The opposite situation occurs in the case of rare earth doped materials, where the $spd$-$f$ hybridisation is usually weak.

Obviously, the way by which the above mechanisms lead to the spatial ordering of the spin polarisations depends on whether the $d$ electrons remain localised on parent ions or undergo an insulator-to-metal transition and, therefore, contribute to the Fermi volume. In the former case, and in the absence of carriers, the spins of magnetic ions are coupled by the mechanism known as the {\em superexchange}. In metals or extrinsic semiconductors, where either conduction or valence $sp$ bands are only partly occupied, a carrier-mediated spin-spin coupling emerges, known as the {\em sp-d Zener} or {\em Ruderman-Kittel-Kasuya-Yosida (RKKY)} mechanism. If doping of a magnetic insulator results in the appearance of electrons in the upper or holes in the lower Hubbard band, the so-called {\em double-exchange} mechanism can operate. Finally, for a sufficiently large overlap between the $d$ wave functions the Mott-Hubbard transition will take place, so that the {\em Stoner-like} mechanism of ferromagnetism in an itinerant carriers systems will dominate. All these mechanisms will be briefly described here below.

\subsection{Superexchange}
As a result of the aforementioned $sp$-$d$ exchange interaction, the valence band electrons will be either attracted to or repulsed by the adjacent magnetic ions, depending on the mutual orientation of the itinerant and localised spins. This results in a spatial redistribution of spin-down and spin-up valence band electrons, the total energy of the system attaining a minimum for an antiferromagnetic arrangement of neighbor localised spins, as shown on Fig.~5. This indirect $d$-$d$ coupling is known as superexchange. Indeed, most of magnetic insulators are antiferromagnets or ferrimagnets (if ions with different spin states co-exist), with N\'eel temperatures reaching 523~K in the case of NiO and Curie temperatures approaching 800~K in ferrimagnetic spinel ferrites (Zn,Ni)Fe$_2$O$_4$.

\begin{figure}[htb]
\begin{center}
\includegraphics[width=0.9\linewidth]{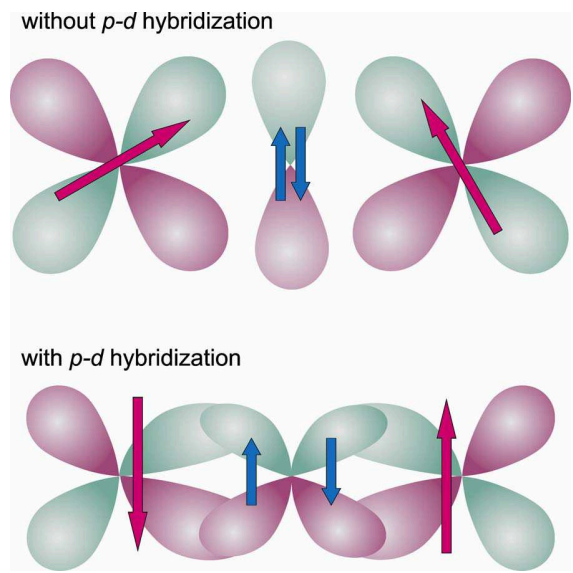}
\caption{Schematic illustration of the antiferromagnetic superexchange between spins (red arrows) localised on open $d$ shells, mediated $\textit{via}$ the $p$-$d$ hybridisation by entirely occupied anion $p$ states. A spin-dependent shift of the orbitals, occurring for an antiferromagnetic arrangement of neighbour Mn spins (lower panel), enhances the lowering of the system energy associated with the $p$-$d$ hybridisation.}
\end{center}
\label{fig:superexchange}
\end{figure}

However, the case of europium chalcogenides (\textit{e.~g.}, EuS) and chromium spinels (\textit{e.~g.}, ZnCr$_2$Se$_4$) implies that the superexchange is not always antiferromagnetic and that ferromagnetism is not always related to the presence of free carriers, even though the Curie temperature $T_{\mbox{\small{C}}}$ does not exceed 100~K in these compounds, despite the large magnetic ion concentration. In the case of rock-salt Eu compounds, there appears to be a competition between antiferromagnetic cation-anion-cation and ferromagnetic cation-cation superexchange \citep{Wachter:1979_a}. The latter can be traced back to the ferromagnetic $s$-$f$ coupling, and the presence of $s$-$f$ hybridisation, that is actually stronger than the $p$-$f$ hybridisation due to symmetry reasons \citep{Wachter:1979_a,Dietl:1994_a}. In such a situation, the lowering of the conduction band associated
with the ferromagnetic order enhances the energy gain due to
hybridisation. The Cr-spinels represent the case, where the $d$
orbitals of the two cations are not coupled to the same $p$ orbital, resulting -- in agreement with the {\em Goodenough-Kanamori rules} -- in a net ferromagnetic superexchange.

\subsection{$sp$-$d$ Zener/RKKY models}
When an $sp$ band is partly filled by carriers but the electrons in the open magnetic shells remain localised, the $sp$-$d$ interaction causes the appearance of spin-polarised carrier clouds around each localised spin. Since the spins of all carriers can assume the same direction if the band is only partly filled, a ferromagnetic ordering can emerge, as noted by Zener in the 1950's in the context of magnetic metals. This ordering can be considered as driven by the lowering of the carriers' energy associated with their redistribution between spin subbands that are split apart in energy by the $sp$-$d$ exchange interaction with localised spins, as shown schematically in Fig.~6 for the case of a $p$-type DMS.  Typically, solid-state effects, such as band anisotropies, spin-orbit interactions, inter- and intra-band exchange couplings have to be carefully taken into account in order to explain the magnitude of $T_{\mathrm{C}}$ and other pertinent characteristics of the system \cite{Dietl:2000_a}.

\begin{figure}[htb]
\begin{center}
\includegraphics[width=1.0\linewidth]{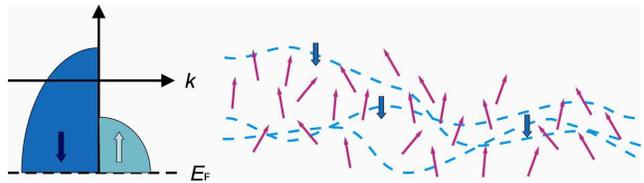}
\caption{Pictorial presentation of carrier-mediated ferromagnetism in $p$-type diluted magnetic semiconductors, a model proposed originally by Zener for metals. Due to the $p$-$d$ exchange interaction, ferromagnetic ordering of localised spins (red arrows) leads to the spin splitting of the valence band. The corresponding redistribution of the carriers between spin subbands lowers energy of the holes (left panel), overcompensates at sufficiently low temperatures an increase of the free energy associated with a decrease of Mn entropy related to the spin ordering.}
\end{center}
\label{fig:Zener}
\end{figure}

A more detailed quantum treatment indicates, however, that the sign of the carrier-mediated interaction between localised spin oscillates with their distance according to the celebrated Ruderman-Kittel-Kasuya-Yosida (RKKY) model.  The RKKY and Zener models lead to the same values of $T_{\mathrm{C}}$ in the mean-field approximation \cite{Dietl:1997_a} as long as the carrier concentration is smaller than that of the localised spins, usually the case of DMSs. Interestingly, the $p$-$d$ Zener model explains the origin of ferromagnetism in double perovskite compounds, like Sr$_2$CrReO$_6$, where the magnitude of $T_{\mathrm{C}}$ attains 625~K \cite{Serrate:2007_a}. If the carrier and spin concentrations are comparable, as in magnetic metals, either ferromagnetic or antiferromagnetic coupling prevails, depending on the ratio of the distance between localised spins to the inverse of the Fermi wave vector. Finally, if the carrier density is much greater that that of the localised spins -- like in diluted magnetic metals -- a random sign of the interaction results in a spin-glass phase.

\subsection{Double exchange}
This mechanism operates if the width $V$ of a partly occupied band is smaller than the energy of the exchange interaction of carriers with localised spins. This situation occurs if magnetic ions with differing charge states coexist. Here, a ferromagnetic spin ordering facilitates the carrier hopping to an empty neighbor $d$ orbital. Thus, since $V$ is proportional to the hopping rate, the ferromagnetic order is driven by a lowering of the carrier energies due to an increase in $V$. Accordingly, in such systems the spin ordering -- either spontaneous or generated by an external magnetic field -- is accompanied by a strong rise of the conductivity, or even by an {\em Anderson-Mott} insulator-to-metal transition,  leading to colossal negative magnetoresistance. This is the case of manganites, like (La,Sr)MnO$_3$, where Sr doping introduces holes in part of the Mn $d$ states.  The ferromagnetic order, surviving up to ~350~K, is brought about by the double-exchange interaction involving the on-site Hund's ferromagnetic spin coupling and hopping of $d$ electrons between neighbor Mn$^{3+}$ and Mn$^{4+}$ ions.

\subsection{Stoner ferromagnetism}
As already mentioned, the potential exchange interaction favors a ferromagnetic ordering of the carriers in a band. However, such an ordering would mean that carriers occupy only one spin subband, so that their Fermi energy, and thus their kinetic energy increases. This increase hampers, at least in a simple one-band case, the appearance of ferromagnetism. However, the existence of the potential exchange leads to a band-gap narrowing in doped semiconductors and enlarges --- by the Stoner enhancement factor or, equivalently, by the Landau's Fermi liquid parameter $A_{\mathrm{F}} > 1$ --- the Pauli susceptibility of normal metals. The same parameter rises the Curie temperature in correspondence to the RKKY/Zener mechanism of localised spins' ordering, $T_{\mathrm{C}} \rightarrow A_{\mathrm{F}}T_{\mathrm{C}}$  \cite{Dietl:1997_a}.

An interesting situation occurs if an overlap between TM $d$ orbitals leads to a {\em Mott-Hubbard} insulator-to-metal transition resulting in the itinerant character of the $d$ electrons, as shown in Fig.~7. If the $d$ band formed in this way is sufficiently narrow, the lowering of the potential energy associated with ferromagnetic ordering is only barely overcompensated by a change in the kinetic energy. In this case the solid-state effects, such as an anisotropic Fermi surface and an $s$-$d$ exchange coupling between the bands, drive the system towards the ferromagnetic phase. This modified Stoner mechanism accounts for the ferromagnetism of, $e.~g.$  elemental ferromagnets, leading to a $T_{\mathrm{C}}$ as high as 1390~K in the case of Cobalt.

\begin{figure}[htb]
\centering
\includegraphics[width=0.7\linewidth]{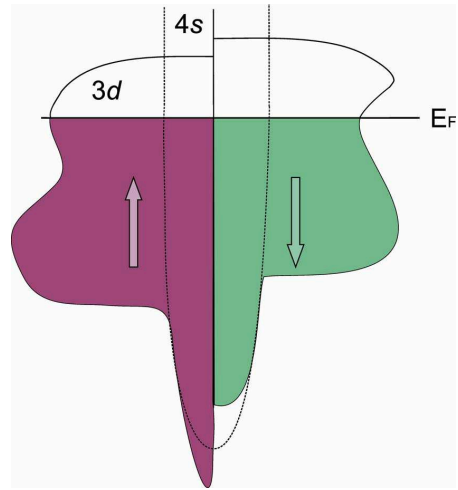}
\caption{Pictorial presentation of the Stoner-like mechanism accounting for ferromagnetism of itinerant magnetic moments. A significant overlap between 3$d$ orbitals hampers the Mott-Hubbard metal-to-insulator transition, so that the 3$d$ electrons are delocalised and form bands. Owing to both the narrowness of these bands and the solid-state effects (Fermi surface anisotropy, $s$-$d$ exchange coupling between bands) the lowering of the Coulomb energy occurring if the electrons are spin polarised overcompensates a corresponding increase in the kinetic energy associated with the redistribution of the electrons to the majority spin subband.}
\label{fig:Stoner}
\end{figure}

\section{Contributions of CMS nanocrystals to ferromagnetic-like features}

{\em A priori} all spin ordering mechanisms specified above can be relevant in the case of CMSs. In particular, a small distance between magnetic ions specific to CMSs can result in the delocalisation of TM $d$ electrons and in the emergence of Stoner ferromagnetism. If the $d$ orbitals retain a localised character, as usually happens in compound ferromagnets, but the material is metallic, a relatively high $T_{\mathrm{C}}$ can result from the carrier-mediated spin-spin coupling, according to the $sp$-$d$ Zener/RKKY model. However, an antiferromagnetic carrier-mediated coupling cannot be ruled out in some cases, according to the RKKY theory. In non-metallic systems, the predominantly antiferromagnetic and strong superexchange, involving entirely occupied anion orbitals, will transfer spin-spin interactions. In cases when magnetic ions with different spin states coexist, the superexchange can lead to a ferrimagnetic spin arrangement.

A high symmetry of the tetrahedral coordination imposed by the host should result in a large density of states at the Fermi level in metallic CMSs, leading to $T_{\mathrm{C}}$ higher  than the one of the free standing counterparts, that typically undergo a global Jahn-Teller distortion to a crystal structure of a lower symmetry.  This expectation is supported by the experimental findings referred to above and showing an enlarged value of $T_{\mathrm{C}}$ for MnAs nanocrystals stabilised in the zinc-blende structure as well as by results of {\em ab initio} computations indicating that a number of zinc-blende transition-metal chalcogenides and pnictides should exhibit a robust ferromagnetism and half-metallicity. By the same token, free standing TM silicides crystallize in low symmetry structures, for which $T_{\mathrm{C}}$ values are low.

Interestingly,  uncompensated spins at the surface of antiferromagnetic nanocrystals can also result in a sizable value of the spontaneous magnetisation up to the usually high N\'eel temperature. It has recently been suggested \cite{Dietl:2008_e} --- but not yet proven --- that the ferromagnetic-like behaviour of (Zn,Co)O in which no precipitates of other crystallographic phases are detected, originates from the presence of wurtzite antiferromagnetic nanocrystals of CoO or Co-rich (Zn,Co)O.  Such nanocrystals can nucleate under appropriate growth conditions or thermal treatment.

We also note that the ferromagnetic proximity effect or the exchange bias in the case of antiferromagnetic CMSs can lead to spin polarisation of a semiconductor surrounding a given nanocrystal. This induced polarisation will persist up to the spin ordering temperature of the CMS and, according to the RKKY theory, will extend over a distance of the order of the inverse Fermi vector in the presence of free carriers and perhaps over two or three bond lengths in their absence.

From the theory of superparamagnetism we know that ferromagnetic or ferrimagnetic nanoparticles exhibit spontaneous magnetisation and magnetic hysteresis up to the blocking temperature $T_{\mathrm{B}} = KV/[k_{\mathrm{B}}\ln(t_{\mathrm{lab}}/\tau)]$, where $K$ is the density of the magnetic anisotropy energy, $V$ is the nanoparticle volume, and $\ln(t_{\mathrm{lab}}/\tau) \approx 25$ for the typical ratio of a relevant spin-flip relaxation time $\tau$ to the time of the hysteresis measurements, $t_{\mathrm{lab}} \approx 100$~s. The value of $T_{\mathrm{B}}$ is further increased by dipolar interactions between nanoparticles, deteriorating also the squareness of the hysteresis loop. The same phenomenology will apply if magnetic moments arise from uncompensated spins in an antiferromagnetic lattice or are generated by the proximity effect or exchange bias.

Quite generally, the magnetic anisotropy has two sources. The first arises from long-range magnetic dipole-dipole interactions, whose energy depends on the orientation of the magnetic moments, i.e., of the magnetisation vector, with respect to axes characterising the shape of the sample.  This shape anisotropy forces the magnetisation vector to lay in-plane or parallel to a long axis in the case of thin films and wires, respectively.  The crystalline anisotropy, in turn, originates from the presence of a spin-orbit interaction within the magnetic ion or within orbitals participating in the exchange process. It typically contains volume and surface or interface contributions, the latter being presumably particularly important for buried CMS nanocrystals. The persistence of ferromagnetic features to high temperatures strongly suggests that the combined effect of shape and crystalline anisotropy results in a magnitude of $K$ sufficiently large to bring $T_{\mathrm{B}}$ close to $T_{\mathrm{C}}$ in the relevant CMS nanocrystals.

\section{Controlling the assembling of nanocrystals}
So far, the most efficient method of controlling the self-organised growth of coherent nanocrystals and quantum dots has exploited the strain fields generated by lattice mismatch at the interfaces of heterostructures \cite{Stangl:2004_a}. Remarkably, in this way it becomes possible to fabricate highly ordered three dimensional dot crystals under suitable spatial strain anisotropy \cite{Stangl:2004_a}.

As already mentioned, the lowering of the growth temperature -- by decreasing the surface and volume diffusion -- hampers the aggregation of magnetic cations and a similar effect can be expected when the growth rate increases. Both these expectations are corroborated by our groups' experimental results for (Ga,Fe)N.

It has recently been suggested that it is possible to change the TM charge state and, therefore, the aggregation energy by co-doping with shallow donors or acceptors \cite{Dietl:2008_e}. This way of affecting the TM valency stems from the presence of band gap states derived from the $d$ orbitals. These states trap carriers supplied by shallow impurities, altering the charge state of the magnetic ions and, hence, modifying their mutual interactions.  Accordingly, co-doping of DMSs with shallow acceptors or donors, during either growth or post-growth processing, modifies $E_{\mathrm{d}}$ and thus provides a mean for the control of the ion aggregation. As an example, the energy of the screened Coulomb interaction between two elementary charges residing on the nearest neighbor cation sites in the GaAs lattice is 280~meV. This value indicates that the Coulomb interaction can preclude the aggregation, as the gain of energy associated with bringing two Mn atoms in (Ga,Mn)As is $E_{\mathrm{d}} = -120$~meV.

It is evident that the model in question should apply to a broad class of DMSs as well as to semiconductors and insulators in which a constituent, dopant, or defect can exist in different charge states under various growth conditions. As important model systems we consider (Ga,Mn)N \cite{Bonanni:2007_b}, (Zn,Cr)Te \cite{Kuroda:2007_a}, and (Ga,Fe)N \cite{Bonanni:2008_a}, where remarkable changes in the ferromagnetic characteristics upon co-doping with shallow impurities have recently been reported. In particular, a strong dependence of the saturation magnetisation $M_{\mathrm{s}}$ at 300~K on co-doping with Si donors and Mg acceptors has been found for (Ga,Mn)N with an average Mn concentration $x_{\mathrm{Mn}} \approx 0.2$\% \cite{Bonanni:2007_b} The model of self-organised growth of nanocrystals explains readily why $M_{\mathrm{s}}$ goes through a maximum when the Mn impurities are in the neutral Mn$^{\mathrm{3+}}$ state, and vanishes if co-doping by the shallow impurities makes all Mn atoms to be electrically charged.

Particularly relevant in this context are data for (Zn,Cr)Te and (Ga,Fe)N, where a strict correlation between co-doping with shallow impurities, magnetic properties, and magnetic ion distribution has been put into evidence. It has been found that the apparent Curie temperature $T_{\mathrm{C}}^{\mathrm{(app)}}$ and the aggregation of Cr-rich nanocrystals depend dramatically on the concentration of shallow Nitrogen acceptors in (Zn,Cr)Te \cite{Kuroda:2007_a}. Actually, $T_{\mathrm{C}}^{\mathrm{(app)}}$ decreases monotonically when the concentration $x_{\mathrm{N}}$ of nitrogen increases, and vanishes when $x_{\mathrm{Cr}}$ and $x_{\mathrm{N}}$ become comparable. This supports the above mentioned model, as in ZnTe the Cr donor level resides about 1~eV above the nitrogen acceptor state. Accordingly, for $x_{\mathrm{N}} \approx x_{\mathrm{Cr}}$ all Cr atoms become ionized and the Coulomb repulsion precludes the nanocrystal formation, as evidenced by Cr spatial mapping \cite{Kuroda:2007_a}.  Importantly, when native acceptors are compensated by Iodine donor doping, both  $T_{\mathrm{C}}^{\mathrm{(app)}}$ and inhomogeneity in the Cr distribution attain a maximum \citep{Kuroda:2007_a}, a behaviour shown in Fig.~8.

\begin{figure}[htb]
\begin{center}
\includegraphics[width=0.7\linewidth]{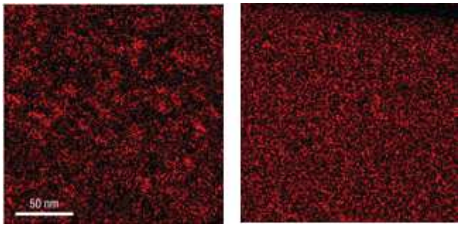}
\caption{Chromium distribution, as obtained from EDS, in (Zn,Cr)Te showing the dependence of spinodal decomposition on co-doping with Iodine (left panel) and Nitrogen (right panel); the marker corresponds to 50 nm. [Adapted by permission from Macmillan Publishers Ltd:  Nature Materials (Kuroda {\em et al.} Nature Mat., 2007, {\bf 6}, 440), copyright (2007)].}
\end{center}
\label{fig:ZnCrTe}
\end{figure}

It has been recently demonstrated that the Fermi-level engineering documented for Iodine and Nitrogen doped (Zn,Cr)Te \cite{Kuroda:2007_a} operates also in the case of III-V magnetic semiconductors. In particular it has been shown that in the model case of wurtzite (Ga,Fe)N the Fermi-level tuning by co-doping with shallow acceptors (Mg) or donors (Si) is instrumental in controlling the magnetic ions aggregation \cite{Bonanni:2008_a}. Specifically, by combining TEM and synchrotron XRD with SQUID magnetometry, three distinct ways by which Fe incorporates into the GaN lattice have been identified, namely: (i) substitutional Fe$^{3+}$ diluted ions accounting for a paramagnetic response of the system \cite{Bonanni:2007_b}; (ii) Fe-rich wurtzite regions of spinodal decomposition commensurate with and stabilized by the GaN host lattice \cite{Bonanni:2008_a} and (iii) hexagonal $\epsilon$-Fe$_3$N secondary phases \cite{Bonanni:2007_b,Bonanni:2008_a}. The  condensation of nanocrystals containing a large density of the magnetic constituent, shown in Fig.~1, clarify the origin of the ferromagnetic features persisting up to above room temperature in this materials system. Moreover, the co-doping with either Si or Mg hampers both the nanocrystal assembling and the onset of spinodal decomposition \cite{Bonanni:2008_a}, providing a striking experimental support to recent theoretical suggestions \cite{Dietl:2008_e}.

Finally, we mention the case of Mn doped GaAs, InAs, GaSb, and InSb. Owing to the relatively shallow character of the Mn acceptors and to a large Bohr radius, the holes reside in the valence band of these systems. Thus, the Mn atoms are negatively charged, a fact that -- according to the model in question -- reduces their clustering, and makes it possible to deposit, by low-temperature epitaxy, an uniform alloy with a magnetic ions content beyond the solubility limit. Co-doping with shallow donors, by reducing the free-carrier screening, will enhance the repulsion among Mn, and allow one to fabricate homogenous layers with even greater $x_{\mathrm{Mn}}$. On the other hand, co-doping by shallow acceptors, along with the donor formation by the self-compensation mechanism will enforce the screening and, hence, lead to nanocrystal aggregation.

\section{Possible functionalities of nanocomposite systems}

The application of embedded metallic and semiconducting nanocrystals is on the way to revolutionise the performance of various commercial devices, such as flash memories, low current semiconductor lasers, and single photon emitters. Similarly far reaching can be the use of nanocomposite semiconductor/ferromagnetic systems due to their unique capabilities and to the possibility of controlling the shape (nanodots {\em vs.} nanocolumns) and size by growth parameters and co-doping during the epitaxial process.

It has already been demonstrated that these nonocomposites show strong magnetotrasport and magnetooptical effects \cite{Kuroda:2007_a,Tanaka:2008_a}, that could possibly allow these systems to be exploited as magnetic field sensors as well as in magnetooptical devices. In particular, a combination of a strong magnetic circular dichroism specific to ferromagnetic metals and weak losses characterising the semiconductor hosts suggest possible functionalities as optical isolators as well as three-dimensional (3D) tunable photonic crystals and spatial light modulators for advanced photonic applications.

As shown in Fig.~9, the controlled growth of nanocolumns of a ferromagnetic metal can allow one to fabricate {\em in-situ}, {\em e.g.}, a dense array of magnetic tunnel junctions \cite{Katayama-Yoshida:2007_a} or Coulomb blockade devices. Thus, the media in question can be employed for low-power high-density magnetic memories, including spin-torque magnetic random access memories and race-track 3D domain-wall based memories.  If sufficiently high tunneling magnetoresistance (TMR) will be found, one can envisage the application for field programmable logic (TMR-based connecting/disconnecting switches) and even for all-magnetic logic, characterised by low power consumption and radiation hardness. Furthermore, embedded metallic nanostructures may serve as building blocks for all-metallic nanoelectronics and for high quality nanocontacts in nanoelectronics,  optoelectronics, and plasmonics as well as constitute media for thermoelectric applications \cite{Katayama-Yoshida:2008_a}. Worth to mention is also the importance of hybrid semiconductor/ferromagnetic systems in various proposals of scalable quantum processors.

\begin{figure}[htb]
\begin{center}
\includegraphics[width=0.9\linewidth]{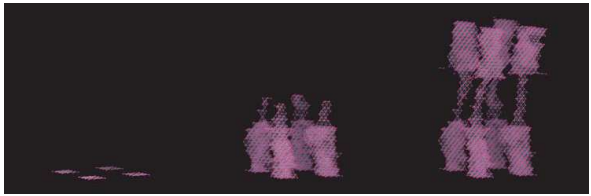}
\caption{Monte Carlo simulation of spinodal decomposition  in (Zn,Cr)Te by two-dimensional spinodal decomposition with seeding and shape control by Cr flux (after \cite{Katayama-Yoshida:2007_a}).}
\end{center}
\label{fig:Hiroshi}
\end{figure}

\section{Mechanisms of exchange interactions between diluted spins}

We now turn to the case of DMSs, in which the employed growth procedure and/or large solubility result in a random distribution of magnetic ions over cation sites. For the $fcc$ or $hcp$ cation sublattice in question, the percolation limit is $x \approx 0.20$ and $ 0.014$ for the nearest neighbour and next nearest neighbour bonds, respectively. Accordingly, for typical TM concentrations, $x \sim 0.05$, the magnetic ions are on average far apart, with a number of important consequences. In particular,  the open magnetic $d$-shells, despite hybridisation with host's bands,  retain a localised character and do not contribute to the Fermi volume, as witnessed by photoemission and $x$-ray spectroscopy, because the concentration of the magnetic constituent is too small for the Hubbard-Mott transition or Anderson-Mott delocalisation to occur within the $d$ states, even if co-doping with shallow impurities results in a co-existence of magnetic ions in two charge states. This means that neither the Stoner-like nor the double exchange mechanisms can account for a long range magnetic order in these systems.  Similarly, the ferromagnetic superexchange even if allowed by symmetry, is expected to be of minor importance in DMSs, as the corresponding magnitude of $T_{\mathrm{C}}$ is already small in CMSs and the relevant exchange integrals decay exponentially with the inter-spin distance.

We are thus lead to the conclusion that a sizable coupling between diluted spins requires the presence of delocalised carriers that can mediate a long-range spin-spin interaction, according to the Zener/RKKY theory.  In agreement with this view, the spin-glass freezing temperature $T_{\mathrm{g}}$ decays much faster with the TM dilution $x$ in diluted magnetic insulators -- like Cd$_{1-x}$Mn$_x$Se, where a short-range antiferromagnetic superexchange dominates and $T_{\mathrm{g}} \sim x^2$, than in diluted magnetic metals -- where typically $T_{\mathrm{g}} \sim x$ according to the RKKY theory. As already mentioned, in the case of extrinsic semiconductors, where the carrier concentration is smaller than the one of localised spins, the carrier-mediated interaction is effectively ferromagnetic. However, for the relevant magnitudes of the density-of-states and the $s$-$d$ exchange integrals, the electron mediated coupling is too weak to overcompensate the antiferromagnetic superexchange and, thus, to produce a long-range ferromagnetic order in DMSs, at least above 1~K \cite{Dietl:1997_a}.

In contrast to $n$-type DMSs, large values of the density-of-states and of the $p$-$d$ exchange integral specific to tetrahedrally coordinated semiconductors result in a strong ferromagnetic spin-spin coupling in $p$-type DMSs, provided that the holes reside in the valence band rather than stay bound to acceptors \cite{Dietl:1997_a,Dietl:2000_a}. Such a ferromagnetism, induced by delocalised or weakly localised holes in Mn-based $p$-type-DMSs, has been observed in III-V compounds \cite{Ohno:1998_a,Matsukura:2008_a}, where Mn ions provide both localised spins and valence band holes, as well as in the case of II-VI systems co-doped by Nitrogen or Phosphorous acceptors \cite{Haury:1997_a,Cibert:2008_a}. Particularly fascinating is the case of Ga$_{1-x}$Mn$_x$As, where the confirmed $T_{\mathrm{C}}$ values reach 180~K for a concentration of the Mn spins randomly distributed over cation sites as low as $x \lesssim 8\%$ \cite{Olejnik:2008_a}.

Comprehensive studies of (Ga,Mn)As and related materials over the last decade have clearly demonstrated a strict relation between the magnitudes of $T_{\mathrm{C}}$ and the hole conductivity. In particular, carrier localisation produced by, for instance, compensation by donors or by isoelectronic disorder in the anion sublattices, has a detrimental effect on the values of $T_{\mathrm{C}}$ and on the spontaneous magnetisation, confirming the notion that the presence of holes in sufficiently extended states is necessary to generate a long-range correlation in the subsystem of localised spins residing in the cation sublattice. At the same time, the $p$-$d$ Zener model \cite{Dietl:2000_a}, in which the spin-resolved structure of the host valence band is carefully parameterised according to the the $kp$ or tight-binding model, describes qualitatively, and often quantitatively with no adjustable parameters, the thermodynamic, micromagnetic, optical, and even magnetotransport properties of hole-controlled ferromagnets  \cite{Dietl:2008_e,Jungwirth:2006_a,Jungwirth:2008_a} as well as of corresponding functional devices, including Zener-Esaki diodes and magnetic tunnel junctions \cite{Dietl:2008_e}. An important aspect of these systems, when compared to earlier studied ferromagnets, is the direct applicability of the time honoured $s(p)$-$d$ model as well as the primary significance of the spin-orbit interaction.

It is important to realize that a sufficiently strong $p$-$d$ hybridisation may turn an isoelectronic TM impurity into a hole trap or enhance significantly the hole ionisation energy, if the TM dopant acts as an acceptor \cite{Dietl:2008_e}. Since the magnitude of the hybridisation increases when the bond length becomes shorter, the effect becomes progressively stronger on going from tellurides to oxides and from antimonides to nitrides. This strong coupling  shifts the insulator-to-metal transition to higher hole concentrations and presumably accounts for the fact that no hole-controlled ferromagnetism has so-far been observed in, for instance, (Ga,Mn)N, where holes remain tightly bound to Mn ions for $x$ at least as large as 6\%.  However, once a sufficiently large hole concentrations will be attained, a high temperature ferromagnetism should appear, owing to the large value of the $p$-$d$ exchange energy in these systems.  It is interesting to recall the case of double perovskite compounds, like Sr$_2$CrReO$_6$, where the $p$-$d$ Zener mechanism leads to magnitudes of $T_{\mathrm{C}}$ as high as 625~K, despite that the distance between localised spins is as large as 0.6 -- 0.7~nm \cite{Serrate:2007_a}, $\textit{i.e.}$ much longer than the separation of 0.5 nm between next nearest neighbor cations in ZnO and GaN. This confirms a potential of the $p$-$d$ Zener mechanism in generating a high-temperature ferromagnetism in TM-doped $p$-type oxides and nitrides \cite{Dietl:2000_a}.

\section{Functionalities of hole-controlled ferromagnetic DMSs}

Due to the fact that ferromagnetic III-V and II-VI compounds can readily be incorporated in semiconductor epitaxial structures, a number of layered and planar devices -- exploiting the unique opportunities offered by these systems -- has been experimentally realized. As recently reviewed \cite{Matsukura:2008_a,Jungwirth:2008_a,Gould:2008_a},  modulated structures containing (Ga,Mn)As show spin transport functionalities relevant for spintronic devices including  spin injection of holes and electrons, anisotropic magnetoresistance (AMR), planar Hall effect,  inter-layer coupling, exchange bias, giant magnetoresistance (GMR), TMR, tunneling anisotropic magnetoresistance (TAMR), Coulomb blockade anisotropic magnetoresistance (CBAMR), and domain wall resistance. As an example, we show in Fig.~10 the degree of electroluminescence circular polarisation, witnessing the high efficiency of electron-spin injection by a Zener-Esaki diode of (Ga,Mn)As.

\begin{figure}[htb]
\includegraphics[width=1.1\linewidth]{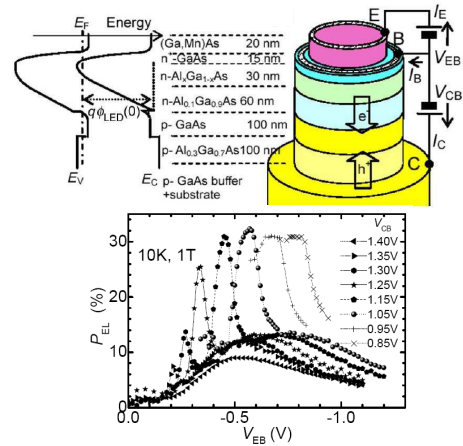}
\caption{Zener-Esaki diode of (Ga,Mn)As integrated with light emitting diode. The high degree of electroluminescence circular polarisation witnesses the efficient electron spin injection into $n$-GaAs. (Reprinted with permission from M. Kohda, T. Kita, Y. Ohno, F. Matsukura, and H. Ohno, Appl. Phys. Lett., 2006, {\bf 89}, 012103. Copyright 2006, American Institute of Physics).}
\label{fig:ZE_diode}
\end{figure}

Since in ferromagnetic DMSs the magnetic properties are controlled by band holes, an appealing possibility is to isothermally influence the magnetic ordering by affecting the carrier concentration in the semiconductor quantum structures by light or by electric field \cite{Haury:1997_a}. Such tuning capabilities of the materials systems in question were found in (In,Mn)As/(Al,Ga)Sb and (Ga,Mn)As as well as in metal-insulator-semiconductor (MIS) devices with a channel of (In,Mn)As or (Ga,Mn)As \cite{Matsukura:2008_a}. The effect of light and  electric field was also put into evidence for modulation doped $p$-(Cd,Mn)Te/(Cd,Mg,Zn)Te:N heterostructures \cite{Cibert:2008_a}, as shown in Fig.~11.  Furthermore, the current-induced magnetisation  reversal was demonstrated in submicron pillars of (Ga,Mn)As/GaAs/(Ga,Mn)As. Electric current was also shown to displace magnetic domain walls in (Ga,Mn)As with the easy axis perpendicular to the film plane \cite{Matsukura:2008_a}.

\begin{figure}[htb]
\begin{center}
\includegraphics[width=1.1\linewidth,clip]{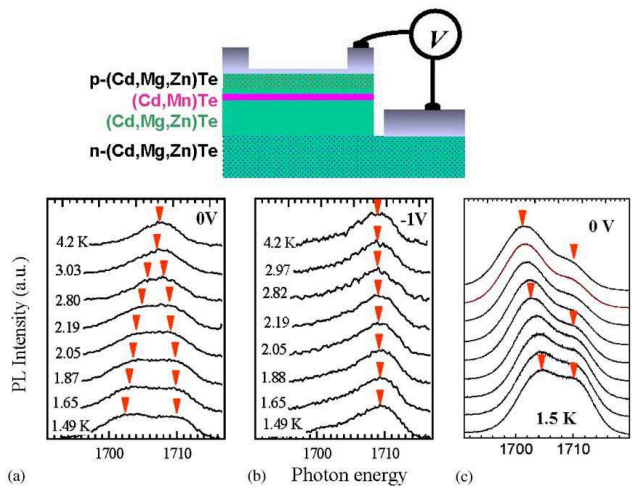}
\caption{Effect of temperature (a), bias voltage (b), and illumination (c) on the photoluminescence of a structure consisting of modulation doped $p$-(Cd,Mn)Te quantum well and $n$-type barrier. The zero-field line splitting (marked by arrows) reveals the appearance of a ferromagnetic ordering (a) which does not show up if the quantum well is depleted from the holes by reverse bias of $p$-$i$-$n$ diode (b). The low-temperature splitting is enhanced by additional illumination with white light (c), that increases the hole concentration in the quantum well (after \cite{Boukari:2002_a}).}
\end{center}
\label{fig:LED}
\end{figure}

These demonstrations of novel functionalities and their successful theoretical understanding have a considerable impact on present research, not only stimulating further works on the development of high-$T_{\mathrm{C}}$ hole-controlled ferromagnetic DMSs, but also prompting the search for similar phenomena in nanostructures of ferromagnetic metals at ambient temperature. Recent works on TAMR, current-induced domain wall displacement, and gating of ferromagnetic metals belong to this stream of ongoing activities.

\section{Is high--$T_{\mathrm{C}}$ ferromagnetism possible without transition metals?}

A major breakthrough in the physics of magnetism as well as in the application of electric engines in -- for instance -- the automotive industry, would be the elaboration of room temperature ferromagnets not containing the heavy transition metals. Organic ferromagnets \citep{Fujiwara:2005_a} and quantum Hall ferromagnets are a proof that ferromagnetism is possible in materials without magnetic ions, albeit the corresponding Curie temperatures are so-far rather low, being to-date settled below 20~K. It has also been suggested that a robust ferromagnetism can appear in certain zinc-blende metals like CaAs, and driven by a Stoner instability in the narrow heavy hole band \cite{Geshi:2005_a}, a prediction awaiting for an experimental confirmation.

Other researchers relate the presence of unexpected high-temperature ferromagnetism in various oxides and carbon derivates to magnetic moments residing rather on nonmagnetic defects or impurities than on open $d$ shells of transition metals \cite{Katayama-Yoshida:2008_a,Coey:2006_a}. It has been known for a long time that a number of defects or non-magnetic impurities form localised paramagnetic centers in various hosts. Some of these centers might show a large intra-center correlation energy $U$ that could ensure an adequate stability of the spins, even if their density increases or the material is co-doped with shallow impurities.

However, similarly to the case of spins residing on TM impurities, it is not easy to invoke a long-range exchange mechanism leading to the high $T_{\mathrm{C}}$ in question, at least if the material does not exhibit any appreciable conductance over the defect or band states. Indeed, in the absence of carriers, exchange interactions between spins of paramagnetic impurities are typically not only short-range but also merely antiferromagnetic \citep{Bhatt:1982_a}. Also, in line with this notion, solid oxygen is an antiferromagnetic insulator.

If, however, the spin concentration increases, so that either the Hubbard-Mott or the Anderson-Mott transition is reached, the Stoner-like mechanism or double exchange might appear. Furthermore, as recently shown \cite{Dietl:2008_e}, a sizable exchange interaction takes place between valence-band holes residing on acceptors and electrons in the conduction band. This demonstrates that band carriers can mediate a Zener-type coupling between spins localised on defect centers. However, in each of these cases a clear correlation between magnetic and transport properties should be visible, analogous to the one observed routinely in manganites, (Ga,Mn)As, and $p$-(Zn,Mn)Te.  In particular, the fabrication of a spintronic structure -- like a magnetic tunnel junction --  working up to high temperatures, would constitute a strong confirmation of the existence of spin transport in these challenging systems.

Alternatively, defects and impurities, similarly to transition metal dopants, could form high spin aggregates in certain hosts. In this case, the presence of ferromagnetic-like features should correlate with the existance of defect agglomerations that can be revealed by employing state-of-the-art nanoanalytic tools.

So-far, however, suggestions concerning defect-related high-temperature ferromagnetism come only from global magnetisation measurements. Therefore, it appears more natural to assume at this stage that a small number of magnetic nanoparticles -- that escaped from the detection procedure -- account for the high-temperature ferromagnetic-like behaviour of nominally nonmagnetic insulators and semiconductors. Such nanoparticles could be introduced during the synthesis or post-growth processing, and can reside in the sample volume, at dislocations or grain boundaries but also at the surface, interface or in the substrate, even if no magnetic response has been found in a cautious magnetic measurement of the unprocessed substrate.

\section{Summary}

In this review, methods employed to fabricate and characterise at the nanoscale tetrahedrally coordinated semiconductors doped with transition metals have been surveyed emphasising the properties, theoretical understanding, and functionalities of these systems. As demonstrated recently, the spatial distribution of magnetic cations can assume various forms in these alloys, opening the door for novel physics and hitherto unrevealed capabilities. While different mechanisms can account for robust ferromagnetic-like features of condensed magnetic semiconductors, delocalised or weakly localised holes in the valence band have been found to be indispensable to promote ferromagnetic coupling between diluted spins. The corresponding $p$-$d$ Zener model explains pertinent properties of hole-controlled ferromagnets and allows for a successful modeling of device structures. It has been argued that the question whether a robust ferromagnetism is possible without transition metals remains open.

\begin{acknowledgments}
The work was supported by FunDMS Advanced Grant of European Research Council within "Ideas" 7th Framework Programme of EC as well as by the Austrian Fonds zur {F\"{o}rderung} der wissenschaftlichen Forschung (P18942, P20065 and N107-NAN).
\end{acknowledgments}

\end{document}